\documentclass[12pt]{article}
\pdfoutput=1
\usepackage{euscript,amsmath,amsthm,amsfonts,amssymb}
\usepackage{mathtools}
\usepackage{subfig}
\usepackage{float}
\usepackage[margin=1in]{geometry}
\usepackage{graphicx}
\usepackage{outlines}
\usepackage[dvipsnames]{xcolor}
\usepackage{jheppub}
\newcommand{\ket}[1]{|{#1}\rangle}

\DeclareMathOperator{\area}{area}

\newtheorem{theorem}{Theorem}
\newtheorem{definition}{Definition}
\newtheorem{conjecture}{Conjecture}
\newtheorem{lemma}{Lemma}

\title{\boldmath Bit threads on hypergraphs}
\author[a,b]{Ning Bao}
\author[c]{and Jonathan Harper}
\affiliation[a]{Computational Science Initiative, Brookhaven National Lab, Upton, NY, 11973, USA}
\affiliation[b]{Center for Theoretical Physics, Department of Physics, University of California, Berkeley, CA
94720, USA}
\affiliation[c]{Martin Fisher School of Physics, Brandeis University, Waltham, Massachusetts 02453, USA}
\preprint{BRX-TH-6671}
\emailAdd{ningbao75@gmail.com}
\emailAdd{jharper@brandeis.edu}
\abstract{Recent work has characterized the various inequalities that entanglement entropies represented by min-cuts on hypergraphs will satisfy. This collection, the hypergraph entropy cone, can be seen as a generalization of the holographic entropy cone which describes the entropies given by both min-cuts on 2-graphs and those of holographic states in AdS/CFT. In this article we describe a generalization of bit threads which allows us to describe max multiflows on hypergraphs. We further comment on its properties and interpretation in holography. }
\begin{document}
\maketitle
\flushbottom


\section{Introduction}

An interesting direction of modern research is the connection between geometry and information. The most famous example of this is the Ryu-Takayanagi (RT) formula \cite{Ryu:2006ef,Hubeny:2007xt} relating the entanglement entropy of a geometric boundary subregion as the area of a homologous minimal (and covariantly, extremal) surface in a holographic spacetime. This allows one to prove many properties of the entanglement entropy simply by examining combinations of different surfaces \cite{Headrick_2007,Hayden_2013}. Two well know examples are subadditivity (SA) and monogamy of mutual information (MMI). SA is given by
\begin{equation}
S(A)+S(B)-S(AB)=I(A:B)\geq 0
\end{equation}
where $I(A:B)$ is the mutual information. This is a universal inequality, meaning that it holds for all quantum states. On the other hand, MMI states that
\begin{equation}\label{eq:mmi}
    -I_{3}(A:B:C)=I(A:BC)-I(A:B)-I(A:C)\geq0
\end{equation}
which can be equivalently written in terms of entropies as 
\begin{equation}\label{eq:mmiS}
    S(AB)+S(AC)+S(BC)\geq S(A)+S(B)+S(C)+S(D),
\end{equation}
where $D$ is the purifier of $ABC$. An important aspect of MMI is that it is not true for all states; rather it is a restriction on the values of the entanglement entropies of various combinations of regions which can appear for holographic states. The procedure of finding all such entropy relations in holography for any number of parties has been formalized as the holographic entropy cone \cite{Bao_2015,Hubeny_2018,Hubeny_2019,Hern_ndez_Cuenca_2019}.

There are however many states which do not satisfy MMI. Consider the four party GHZ state, an important example of true 4-party entanglement (see e.g. \cite{bengtsson2016brief})
\begin{equation}
    \ket{GHZ_4}=\frac{1}{\sqrt{2}}(\ket{0000}+\ket{1111}).
\end{equation}
For this state all of the entanglement entropies have the same value $S(A_i)=S(A_iA_j)=1$ and thus has $-I_3=-1$. Thus, this state violates MMI and implies that no holographic state can recreate such GHZ-like entanglement at leading order. This leads naturally to the question of if this and other such states can be represented geometrically. The answer to this is yes, with hypergraphs being the geometrical system that implements this. A hypergraph is a collection of vertices and hyperedges with the property that a hyperedge can associate any number of vertices. We often refer to the valence of a hypergraph as the highest number of vertices a hyperedge has, e.g. a $k$-graph has maximal hyperedge valence $k$. Thus, ordinary graphs are simply 2-graphs in this labeling.

\begin{figure}[H]
\centering
\includegraphics[width=.8\textwidth,page=1]{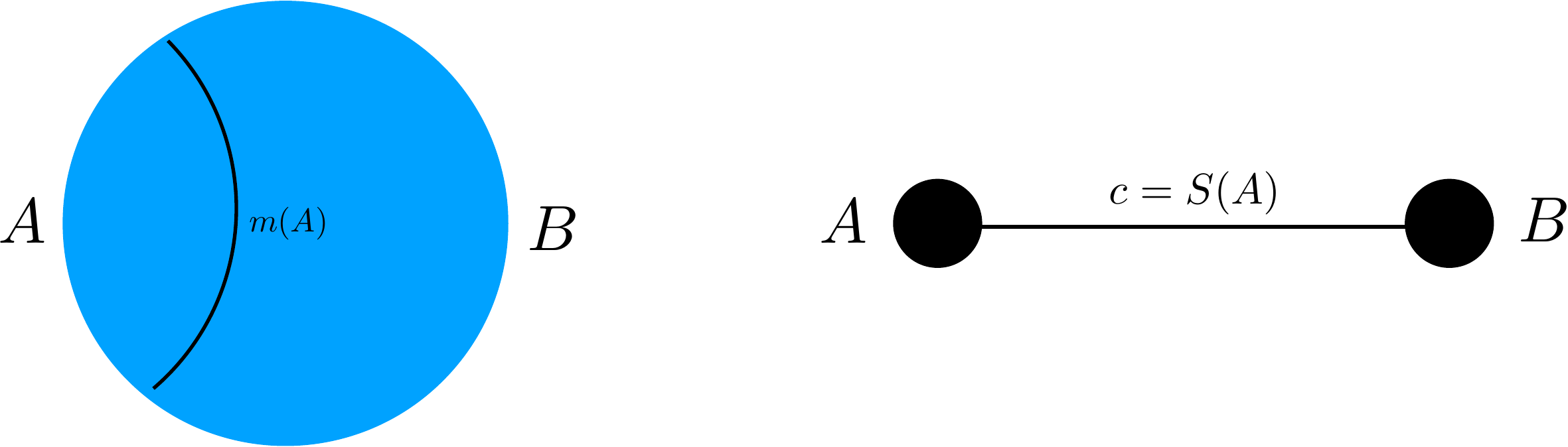}
\caption{\label{fig:AdStoGraph} Starting with a pure state of a constant time slice of AdS we can calculate the entanglement entropy using the RT formula as the area of the minimal area surface homologous to $A$. This picture can be schematically represented as a simple graph of two vertices connected by a single edge with weight equal to $S(A)$. This procedure can be generalized to more regions and more complicated geometries which allow us to consider systems with a particular values for the various entanglement entropies.}
\end{figure}

2-graphs have a special connection to holographic systems. Often we are only interested in the values of the different entropies for all possible combinations of regions. We can then simplify the spacetime by representing this information as a graph. To do so we choose a collection of non-overlapping RT surfaces. For each boundary region we have a boundary vertex and for each internal region we have an internal vertex. These regions are defined as the bulk regions that are completely bounded by potions of RT surfaces and are not further partitioned by additional RT surfaces. For each RT surface we draw an edge with capacity equal to the area of the RT surface (see figure \ref{fig:AdStoGraph}). By this reduction, it is clear that 2-graphs share the same holographic entropy cone. With this in mind, hypergraphs are a natural generalization, but in order to get non-holographic states we must do something different from a simple RT-based discretization of a Riemannian geometry. From the hypergraph perspective this is the inclusion of higher valence hyperedges.

Similarly to the holographic entropy cone, the various entropy inequalities that hypergraphs satisfy have been characterized as the hypergraph entropy cone \cite{Bao_2020,Bao:2020mqq,walter2020hypergraph,Linden:2013kal}. An immediate and key difference is that for four parties one does not have MMI as a valid inequality for all hypergraphs. Instead one must consider the weaker Ingleton's inequality on a 5-party pure state
\begin{equation}
    I(A:B|C)+I(A:B|D)+I(C:D)-I(A:B)\geq0
\end{equation}
or expanded in terms of entropies
\begin{equation}
   S(AB)+ S(AC)+S(AD)+S(BC)+S(BD) \geq S(A)+S(B)+S(CD)+S(ABC)+S(ABD).
\end{equation}
Ingleton's holds for any choice of entropies that can be represented as a hypergraph such as our GHZ state.

The purpose of this paper is to provide an alternate interpretation of hypergraphs and to understand why MMI fails, but Ingleton's inequality is valid. To do so we make use of the bit thread formalism \cite{Freedman:2016zud,Headrick:2017ucz} which allows one to calculate entropies as the maximal flow through the geometry one can create subject to constraints. In proving MMI one must consider max multiflows (MMFs) which have the added condition of saturating all of the single party entropies. In this construction one thinks of a decomposition of the flow which groups together threads that connect the same boundary regions.

For hypergraphs a MMF may not exist so it is necessary to define instead a hypergraph max multiflow (HMMF). The main difference between the MMF and HMMF lies in how the flow interacts with hyperedges. A flow crossing a hyperedge can be simultaneously sent to any connected hyperedges. This causes threads to split forming an ``entanglement web". Importantly, this allows $k$-threads to exist in a HMMF which connect $k$ different regions. For MMI we will see that the existence of such 4-threads directly leads to a violation.

The organization of this paper will be as follows. In section \ref{sec:2} we review the basic definitions and results for bit threads and hypergraphs. Those already familiar with these concepts can safely skip ahead. Next, in section \ref{sec:3} we describe our framework for hypergraph max multiflows, prove their existence and show how they can be used to understand why MMI is violated. We prove a weaker version of Ingleton's inequality which holds when the flow is able to saturate the 2-party entropies and provide a conjecture on how this result might be strengthened. We also provide some comments on possible interpretations of hypercuts and multipartite entanglement in holography.

\section{Bit threads and hypergraphs}\label{sec:2}
\subsection{Bit threads}
Here we review the bit thread formulation for the calculation of entanglement entropy in holographic spacetimes. Starting from a static time slice we first choose a boundary region $A$. The Ryu-Takayanagi (RT) formula \cite{Ryu:2006ef,Hubeny:2007xt} then states that the entanglement entropy can be calculated as the area of the minimal surface homologous to $A$ 
\begin{equation}
S(A) = \min_{m\sim A} \area(m).
\end{equation}
Using techniques from convex optimization the entanglement entropy can alternatively be calculated as
\begin{equation}
S(A)= \max_{v^{\mu}} \int_{A} \sqrt{h} n_{\mu}v^{\mu}\quad\text{s.t. }\nabla_\mu v^\mu=0,\;|v^\mu|\leq 1.
\end{equation}
This can be thought of as the flow of information or ``bit threads" \cite{Freedman:2016zud,Headrick:2017ucz} from $A$ to its complement. The first constraint ensures that threads can only be created or sourced from the boundary while the second places an upper bound on the flux of the flow at any point in the manifold. The RT surfaces act as a bottleneck limiting the number of threads to be exactly the entanglement entropy.

When we have multiple boundary regions we can choose to instead form a max multiflow (MMF) \cite{Cui:2018dyq}. This flow has the added property that the flow will saturate on each RT surface whose area is the single party entropy $S(A_i)$. More explicitly a multiflow is a collection of vector fields $\mathcal{N}_{m} = \{N_{ij}^{\mu}\}$ which satisfy the following properties:
\begin{outline}
\1$n_{\mu}N^{\mu}_{ij} =0 \; \text{on } A_{k}, \; k \neq i,j$
\1 $N_{ij} = -N_{ij}$
\1 $\nabla_{\mu}N_{ij}^{\mu} = 0$
\1$\sum\limits_{i<j}\left|N_{ij}\right| \leq 1.$
\end{outline}
One should think of $N_{ij}$ as the collection of threads which connect $i$ to $j$. The constraints guarantee that everywhere in the manifold this collection of vector fields can be thought of as a single flow - nowhere is the norm bound violated.

We now define a subflow which is essentially the collection of $N$'s which are sourced from a particular boundary.
\begin{definition}[Subflow]
Given a multiflow $\mathcal{N}_{m}$ consider the union of any number of boundary regions in $\mathcal{A}$ denoted $B$. The subflow of $B$ is the sum of all flows whose threads connect to $B$
\begin{equation}\label{eq:subflow}
V_{B} = \sum\limits_{i,j \text{ s.t } A_{i} \in B, \; A_{j} \notin B} N_{ij}.
\end{equation}
\end{definition}
\noindent From this we can state the following theorem:
\begin{theorem}[Existence of a maximal multiflow]\label{existmultiflow}
There exists a multiflow $\mathcal{N}_{m}$ such that all single interval subflows are maximal. That is
\begin{equation}
S(A_{i}) = \int_{A_{i}} V_{A_{i}}
\end{equation}
for some $\mathcal{N}_{m}$. In general the subflows associate to higher party regions will not be maximal.
\end{theorem}
\noindent From now on we will instead write $V_A$ ($N_{Aj}$) to mean ``the flux of $V_A$ ($N_{Aj}$) on $A$" leaving the integration implicit. This is done to more closely mirror the notation used on graphs and hypergraphs. 

For a 4-party pure state with regions $A,B,C,D$ we can use Theorem \ref{existmultiflow} to guarantee that
\begin{equation}\label{eq:1region}
    S(A)=V_{A}=N_{AB} +N_{AC} + N_{AD},
\end{equation}
however for any 2-party region e.g. $AB$ we only have
\begin{equation}\label{eq:2region}
    S(AB)\geq V_{AB}=N_{AC} + N_{AD}+N_{BC} +N_{BD}.
\end{equation}
Note that here $N_{AB}$ does not contribute since these threads are internal between $A$ and $B$ and not from $AB$ to the other boundary regions. This decomposition allows one to prove MMI using bit threads:
\begin{theorem}[Monogamy of mutual information (MMI)] Consider the case of four boundary regions $A,B,C,D$
\begin{equation} \label{MMI}
-I_{3}(A:B:C) = S(AB)+S(AC)+S(BC) -S(A)-S(B)-S(C)-S(ABC) \geq 0
\end{equation}
\end{theorem}
\noindent The proof follows almost immediately by writing out the subflows and using existence of a max multiflow. Consider the two interval subflows $V_{AB},V_{AC},V_{BC}$. In general these can not all be made maximal, thus
\begin{equation}
\begin{split}
 &S(AB)+S(BC)+S(AC) \geq V_{AB} +V_{AC} + V_{BC} \\   
&=(N_{AC}+N_{AD}+N_{BC}+N_{BD})+(N_{AB}+N_{AD}+N_{BC}+N_{CD})+(N_{AB}+N_{AD}+N_{AC}+N_{AD})\\
&=(N_{AB}+N_{AC}+N_{AD})+(N_{AB}+N_{BC}+N_{CD})+(N_{AC}+N_{BC}+N_{CD})+(N_{AD}+N_{BD}+N_{CD})\\
&=  V_{A} +  V_{B} + V_{C} + V_{D}
\\ &= S(A) +S(B) + S(C) + S(ABC).
\end{split}
\end{equation}
In the first two lines we used \eqref{eq:2region}, in the third we rearranged terms, and in the last we used \eqref{eq:1region} and purity of the state. Finally, subtracting the two sides gives MMI.

It is important to remember going forward that the key feature of this proof is the ability to saturate simultaneously all of the single party RT surfaces. These results all generalize straightforwardly to 2-graphs.

\subsection{Hypergraphs}
Here we define a hypergraph and an oriented hypergraph. For a given hypergraph many oriented hypergraphs can be constructed. Of these oriented hypergraphs we define the notion of a valid orientation. For our purposes this will be the subset of all orientations which can support flows.

\begin{definition}[Hypergraph]
A hypergraph is a set of vertices $X$, a set of hyperedges $E$, along with a capacity vector $C^e$ which assigns to each hyperedge a positive number. From these we can construct the incidence matrix $I_{e,x}$ where each element is 1 if the vertex belongs to that hyperedge and 0 otherwise. Each row can have multiple 1's.
\end{definition}

\begin{figure}[H]
\centering
\begin{tabular}{cc}
\includegraphics[width=.45\textwidth,page=2]{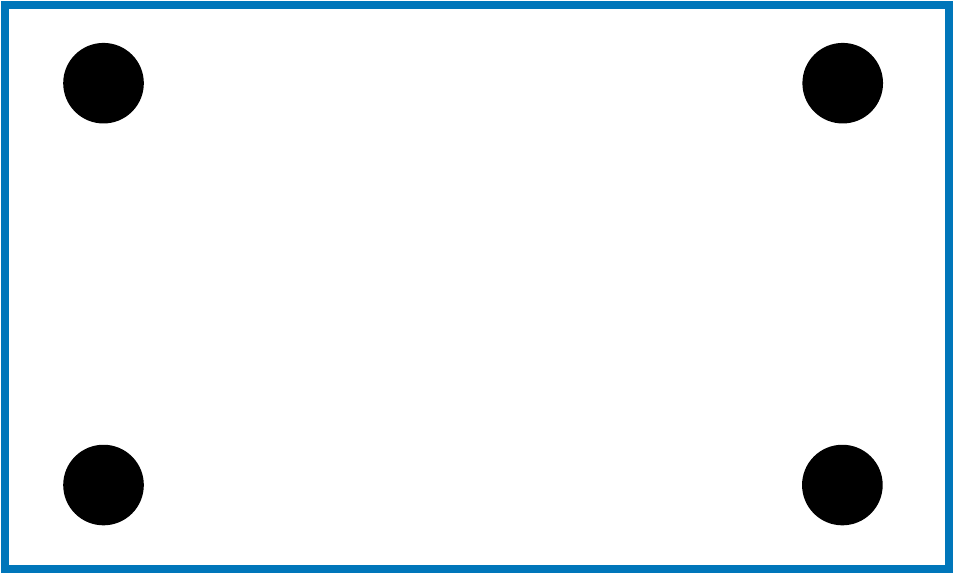} & \includegraphics[width=.45\textwidth,page=3]{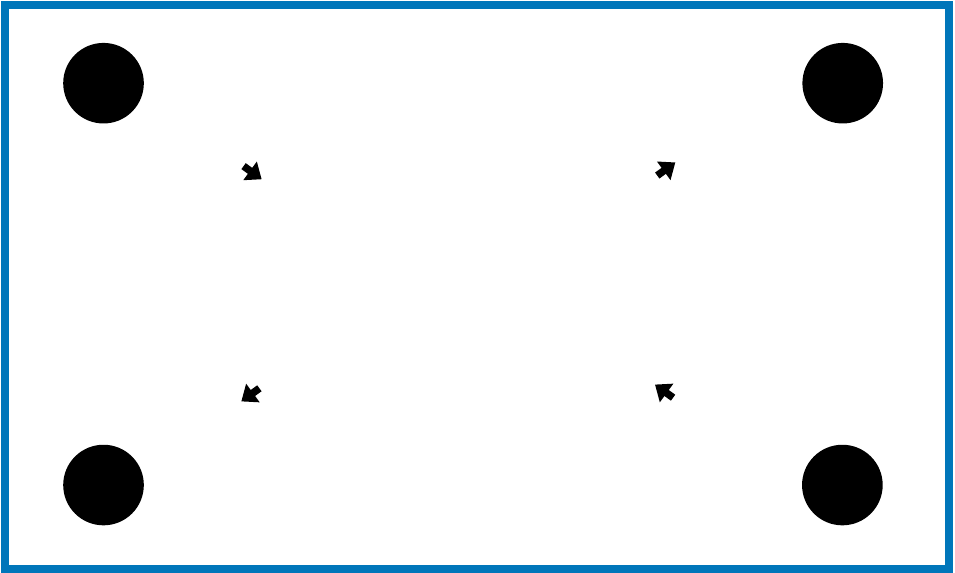} 
\end{tabular}
\caption{\label{fig:GHZHG} Left: A simple hypergraph with 4 vertices $X=\{A,B,C,D\}$ a single 4-edge $e_1$ with capacity $C^{e_1}=1$ and incidence matrix $I_{e_{1},x}=\{1,1,1,1\}$. Right: The same hypergraph with a valid orientation $I_{e_{1},x}=\{-1,1,-1,1\}$}
\end{figure}

\begin{definition}[Oriented Hypergraph]
An oriented hypergraph is a set of vertices $X$, a set of hyperedges $E$, along with a capacity vector $C^e$ which assigns a positive number to each hyperedge. From these we can construct the incidence matrix $I_{e,x}$ where each element is 1 (-1) if the vertex belongs to that hyperedge and is oriented out from (in to) the vertex, otherwise it is 0.
\end{definition}
\begin{figure}[H]
    \centering
    \includegraphics[width=.5\textwidth,page=4]{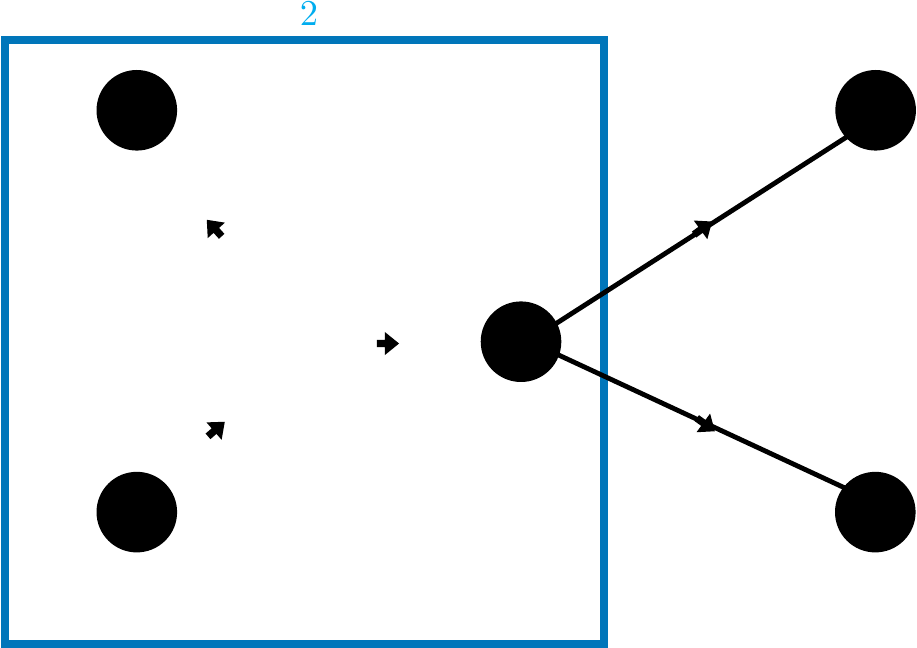}
    \caption{\label{fig:orientation} Another example of an oriented hypergraph. Note that all hyperedges and the internal vertex have at least one incoming and outgoing arrow.}
\end{figure}
\noindent We will call an orientation \emph{valid} if every row and each column associated to an internal vertex contains at least one $1$ and $-1$ (see figures \ref{fig:GHZHG} and \ref{fig:orientation}).

\section{Hypergraph multiflows}\label{sec:3}
In order to discuss entropies relating more than two boundary regions, it will be necessary for us to make use of a suitable generalization of a multiflow. We wish to maintain the key properties we have in the 2-graph case: flow conservation at vertices, bounding by hyperedge capacities, and the saturation of the the single party entropies.\footnote{In the literature, definitions of flows on hypergraphs exist, however they are more akin to a simple generalization of 2-graph flows in that each thread can only connect two vertices \cite{menger}. The need for our definition arises because of the requirement for all of the single party entropies to be simultaneously saturated.}

\begin{figure}[H]
\centering
\includegraphics[width=.4\textwidth,page=5]{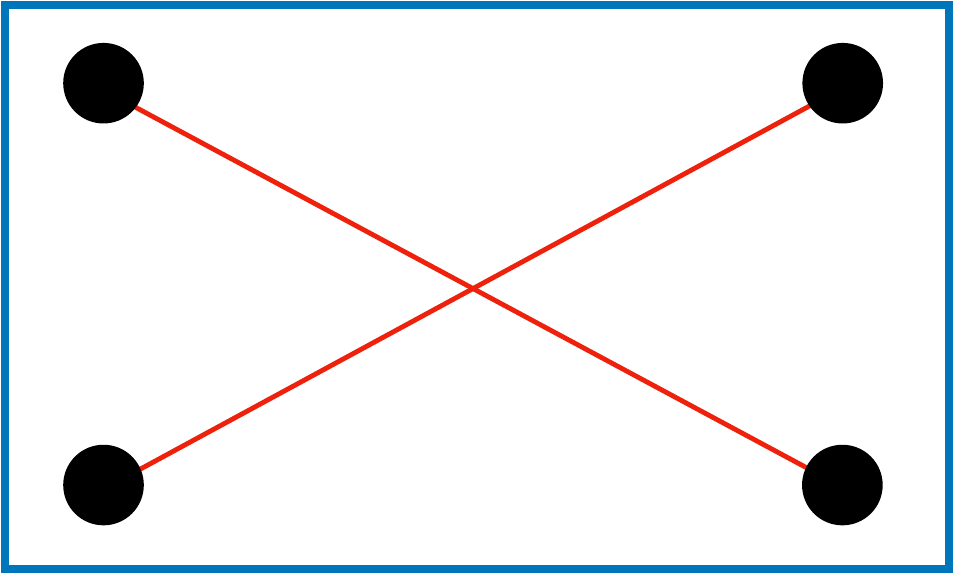}
\caption{\label{fig:GHZ} A hypergraph which consists of a single 4-edge, shown in blue. This hypergraph has the same entanglement entropies as the 4-party GHZ state and thus acts as an avatar of true 4-party entanglement. The HMMF is given by a single thread which explicitly and simultaneously connects all four boundary regions.}
\end{figure}

To start we can consider some simple hypergraphs such as the GHZ$_{4}$ state formed by a single 4-edge (see figure \ref{fig:GHZ}). Since the hyperedge has capacity $1$ we can only have one thread, but for the single party entropies to be saturated, that thread must connect each boundary vertex. We are immediately led to the conclusion that our definition of a hypergraph max multiflow must allow threads which can simultaneously connect multiple boundary regions. To this end we make the following changes to the constraints:

\begin{figure}[H]
\centering
\begin{tabular}{cc}
\includegraphics[width=.35\textwidth,page=6]{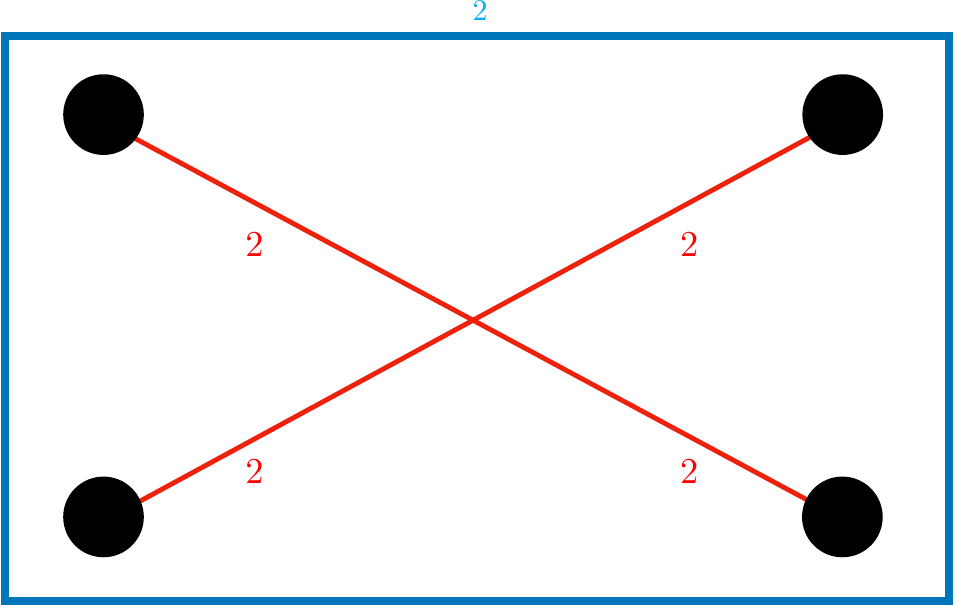} & \includegraphics[width=.35\textwidth,page=7]{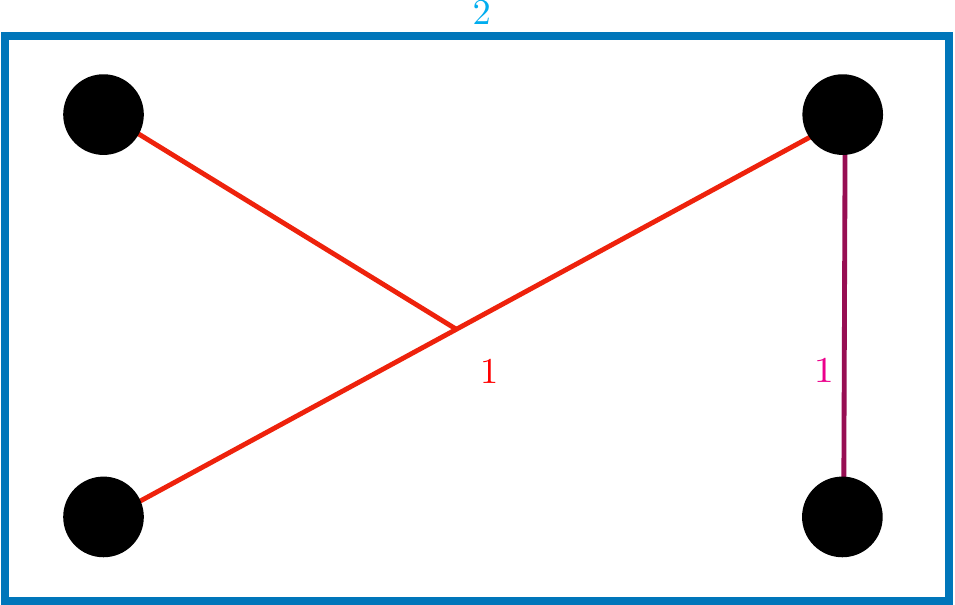} \\
\multicolumn{2}{c}{\includegraphics[width=.35\textwidth,page=8]{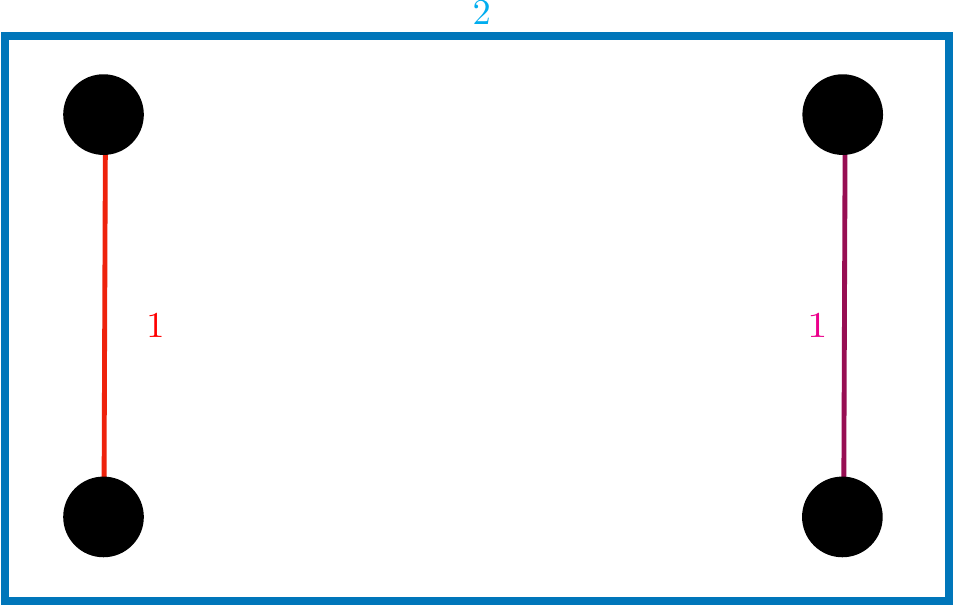} }
\end{tabular}
\caption{\label{fig:cap} The following thread configurations all saturate the capacity constraint of 2 on the 4-edge. Each ``use" of the hyperedge places a thread connecting any number of vertices and consumes one capacity. Each placement of a thread can be represent using fluxes by increasing the magnitude of the flux of each hyperedge-vertex pair the thread connects by 1.}
\end{figure}

\paragraph{Capacity constraint}
In order to define a HMMF, it will be necessary to talk about fluxes on a hypergraph. However, the easiest way to understand the capacity constraint is by thinking about placing individual threads on the hypergraph and considering their arrangement on the hyperedge of interest. When a thread travels through a hyperedge, it has the freedom to simultaneously travel to any or all vertices the hyperedge connects. This action requires a single unit of capacity. Thus, the total number of threads a hyperedge can support is exactly the capacity. Note that independent threads need not connect to the same vertices (see figure \ref{fig:cap}).

To describe the flux we first define the matrix $V_{e,x}$ each entry of which can be thought of as particular hyperedge-vertex pair ($V_{e_1,x_1}$ e.g.). Each entry is given a value equal to the total number of threads connecting to $A_i$ through $e_i$. For an oriented hypergraph, the sign of each entry is given by the sign of the corresponding element of the incidence matrix $I_{e,x}$. The capacity constraint is then the requirement
\begin{equation}
    |V_{e,x}|\leq C_e, \quad \forall\; e,x.
\end{equation}

One may worry that this constraint is too weak because flux is not conserved on the edges. We will show that it is always possible to find a HMMF for which the incoming and outgoing flux is the same on each edge\footnote{This is automatically true for 2-edges.}. For odd-threads this requires additional attention: an auxiliary boundary vertex $Z$ must be added to guarantee the flux is balanced. Once a flux configuration for a HMMF is found, the contributions $V_{e_i,Z}$ can be set to zero and the threads resolved on the original hypergraph.

\begin{figure}[H]
\centering
\begin{tabular}{cc}
\includegraphics[width=.4\textwidth,page=9]{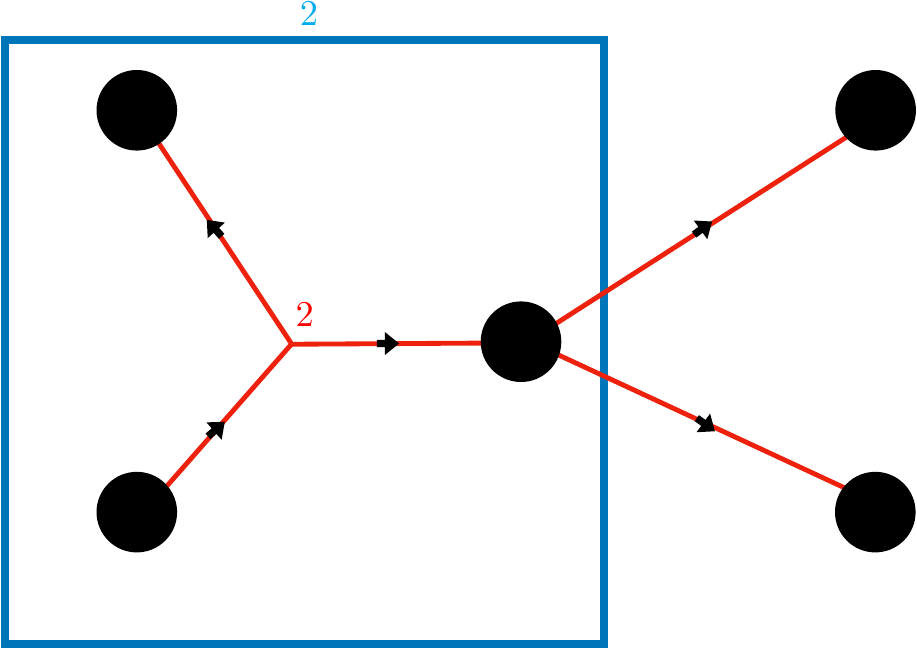} & \includegraphics[width=.4\textwidth,page=10]{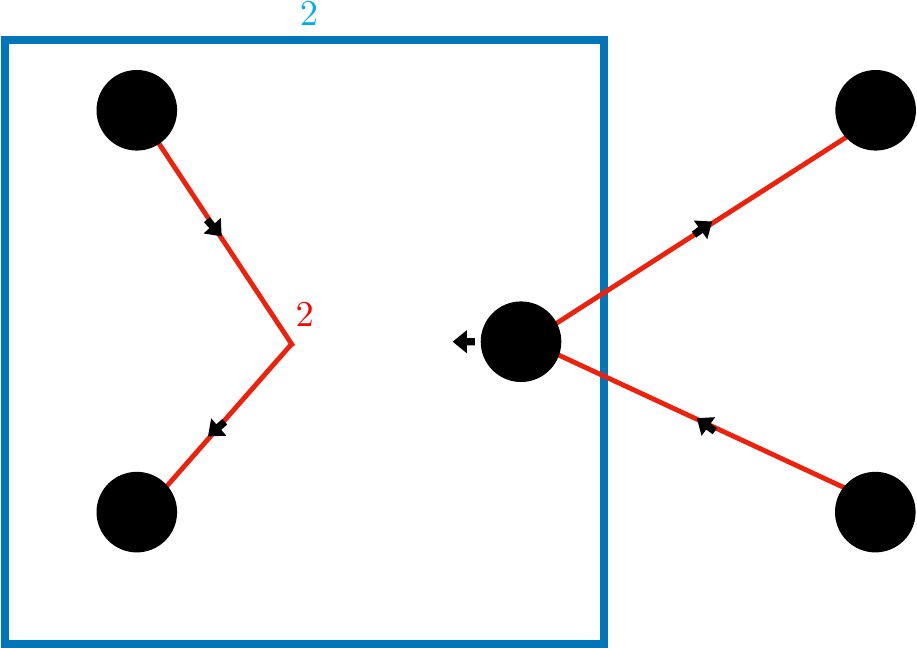} 
\end{tabular}
\caption{\label{fig:sumconstraint}When looking for a HMMF we start with an unoriented hypergraph and then consider all valid orientations of that graph. On each we maximize the flux on the graph subject to the constraints. The specified orientation allows us to implement the flow conservation constraint. Not all choices of orientation will produce HMMF. Shown are two different choices of orientation and respective for a particular hypergraph. In this case both are valid HMMFs. Once a valid HMMF is found, one can discard the orientation and view the HMMF as a collection of threads on the original unoriented hypergraph.}
\end{figure}

\paragraph{Flow conservation}
Once an orientation has been chosen, flow conservation on the internal vertices remains essentially unchanged (see figure \ref{fig:sumconstraint}). We require that the incoming and outgoing flux, as defined by a valid orientation, must be the same, that is
\begin{equation}\label{eq:flowcos}
\sum_e V_{e,x_i}=0 \quad \forall x_{i}\notin\partial X.  
\end{equation}
Once the fluxes have been resolved into threads, \eqref{eq:flowcos} is the requirement that threads cannot end on internal vertices, only boundary vertices.

\vspace{\baselineskip}
\noindent We will now define a hypergraph multiflow and a HMMF:
\begin{definition}[Hypergraph multiflow]
For a given hypergraph, a hypergraph multiflow is an assignment to each hyperedge-vertex pair a positive number using the following procedure:

\vspace{\baselineskip}
\noindent First, for or each valid orientation of the hypergraph we assign to each hyperedge-vertex pair a positive or negative number such that:
\begin{outline}
\1 $|V_{e,x}|\leq C_e, \quad \forall\; e,x$.
\1 $\sum_e V_{e,x_i}=0 \quad \forall x_{i}\notin\partial X$.
\end{outline}
This assignment is in general not unique. Next, for a given multiflow we can take the absolute values of each element to define a hypergraph multiflow for the original unoriented hypergraph.
\end{definition}

\begin{definition}[Hypergraph max multiflow (HMMF)]
A HMMF is a hypergraph multiflow such that
\begin{equation}
    S(A_{i})=|\sum_{e} V_{e,A_i}|,\quad \forall A_i \in\partial X.
\end{equation}
\end{definition}
\noindent For our HMMF, this collection of hyperedge-vertex pairs defines our subflows. That is,
\begin{equation}
   V_{A_i}\equiv |\sum_{e} V_{e,A_i}|.
\end{equation}
Each of these subflows can also be resolved into thread numbers. Because threads can connect multiple boundary vertices, this introduces a number of extra terms into the decomposition. For a valence $k$ hypergraph with $n$ boundary vertices we can expect to have 2...$n$-threads contributing to the entropies $N_{AB},...,N_{ABC},...,N_{ABCD}$,... etc. The number of terms contributing to the $l$ threads is given by the number of combinations $\prescript{}{n}{C}_l$.

When calculating a subflow in terms of threads one must be careful to include the correct terms. Any thread combination that contributes must have at least one index included in and excluded from the specified regions. For example, if we have 4-party state with boundary regions $A,B,C,D$ and we want to determine $V_{ABC}$ we do not include the terms $N_{AB}.N_{AC},N_{BC}$ or $N_{ABC}$. The decomposition becomes
\begin{equation}
    V_{ABC}=N_{AD}+N_{BD}+N_{CD}+N_{ABD}+N_{ACD}+N_{BCD}+N_{ABCD}.
\end{equation}
This is the same as $V_D$ as one would expect from purity of the state. More generally, 
\begin{equation}
  V_{A_i...A_k}=V_{(A_i...A_k)^c}  
\end{equation}
where $c$ is the complement.

\subsection{Existence}

Our goal is to prove the existence of a HMMF for an arbitrary hypergraph. We will accomplish this by making use of Theorem \ref{existmultiflow}. First we observe that given any 2-graph we can freely make the following choices:
\begin{outline}
\1 We can define a state with $n$ boundary regions for any hypergraph by partitioning the boundary vertices into $n$ groups. These can then be combined by identifying all members as a single boundary vertex to form $A_i...A_n\in \partial X$
\1 If a boundary vertex is connected to multiple edges, we can always make the edge an internal vertex. To do this we connect it to a new boundary vertex using a single 2-edge with capacity equal to the single party entropy (which can be ascertained from the MMF or by the min-cuts on the graph). This is important as it means we can think of the bottleneck to the single party flows as being on these 2-edges and not in the internal region of the graph.
\end{outline}
Next we organize the graphs into equivalence classes labeled by the values of single party entropies $S(A_i)$. In each of these equivalence classes exist graphs which only contain 2-edges, so by Theorem \ref{existmultiflow} these graphs have a max multiflow which simultaneously saturates the single party entropies.

What is left to be shown is that promoting a collection of hyperedges to a higher valence edge can only change to higher party entropies so that we remain in the same equivalence class. Intuitively, the flows on hypergraphs have much more freedom so a HMMF must exist.

Choosing a particular set of single party entropies, we can always find a 2-graph with the matching values. Because this is a 2-graph we know there exists a MMF. Now we can consider two operations: to this graph we can add or remove $k$-edges. Adding a hyperedge cannot affect the existence of the MMF because the original path of edges can still be used. Thus we are free to add hyperedges of any valence to the 2-graph. We can also delete hyperedges. If deleting a hyperedge would change the single party entropies, then the resulting graph would lie in a different equivalence class. So we can consider just the class of hyperedge deletion operations which leave the single party entropies unchanged. What remains to show is that such a deletion cannot prevent the existence of a HMMF. This relies essentially on the idea that hyperedges are more flexible in the arrangements of flow they can support. Thus, we have the ability to move multiple flows to a single hyperedge. That is, these higher valence edges are ``Pareto efficient" in that a single edge provide the flow with many more allowed configurations. Since the single party entropies are unchanged and the hyperedges allow more freedom and efficiency there must exist a HMMF. Our proof will therefore be constructive, using the MMF to create a valid HMMF. An important case for these operations, which facilitates a proof, is that of adding a single hyperedge and deleting all other hyperedges. 

\begin{figure}[H]
\centering
\begin{tabular}{cc}
\includegraphics[width=.45\textwidth,page=11]{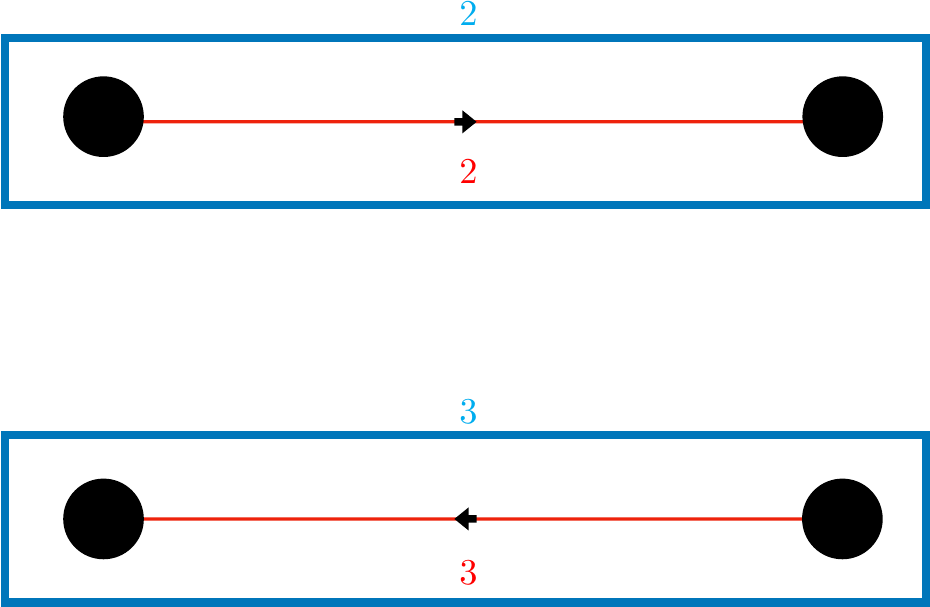} & \includegraphics[width=.45\textwidth,page=12]{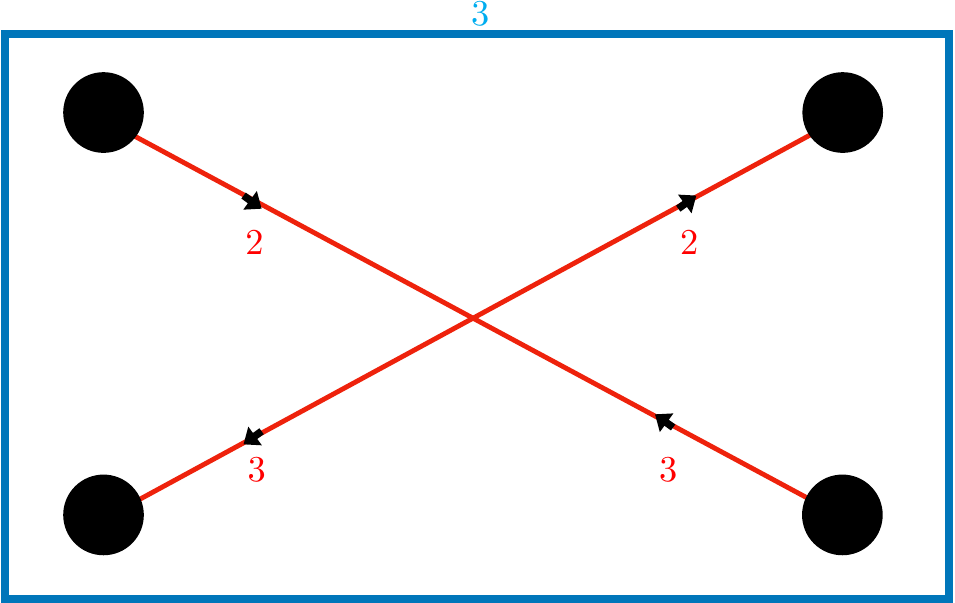}
\end{tabular}
\caption{\label{fig:proof} We start with a 2-graph which has a valid max multiflow. On this graph we pick any choice of subgraph. Left: A particular subgraph with two 2-edges of different capacities. Not shown is the flux on the boundary vertices of the subgraph into the full graph. Right: The two edges have been replaced by a single 4-edge with capacity 3. The orientation and flows follow from the original 2-graph. This procedure is such that the total flux on each boundary vertex is the same. This means if in the full original graph we use the new subgraph, which uses hyperedges, we have a valid HMMF. This procedure can be repeated, allowing us to produce a HMMF on any even hypergraph we wish that is in the same equivalence class as the original 2-graph.}
\end{figure}

To be specific, let us suppose we have an oriented 2-graph and a max multiflow on it. We examine an arbitrary subgraph and combine all of the 2-edges into a single $k$-edge where $k$ is the number of boundary vertices. Using the orientation for the 2-edges, we sum the total flux leaving a boundary vertex and assign the hyperedge-vertex pair of the incidence matrix $-1,0,1$ depending on the result. This provides the necessary orientation for the hypergraph which includes this new $k$-edge. The capacity of the new $k$-edge is assigned an arbitrary value, so long as it is large enough not to change the equivalence class of the hypergraph. This is easily accomplished by picking the capacity to be the largest magnitude of the flux among the newly created edge-vertex pairs. The new HMMF can then be assigned the values of the sum of fluxes of the original 2-graph. This assignment ensures that the conservation constraints on each internal vertex (including the boundary vertices of our subgraph) remain valid. Importantly, since the flow was only changed on the subgraph the bottleneck for the single party entropies remains on the boundary 2-edges. This procedure can be straightforwardly generalized to subgraphs with hyperedges. 

A feature of our construction is that it will always produce a HMMF such that, on each edge, the incoming and outgoing flux will be equal. Namely, this means we will only produce even-threads on a hypergraph. This is only possible if the odd-edges produced can be decomposed into even-edges. We will refer to a hypergraph with only even-edges as an even hypergraph.

Thus, multiple iterations provide us with a constructive process by which to generate a valid HMMF for any even hypergraph given any choice of starting equivalence class. We summarize this process in figure \ref{fig:proof}.

\begin{figure}[H]
\centering
\includegraphics[width=.5\textwidth,page=18]{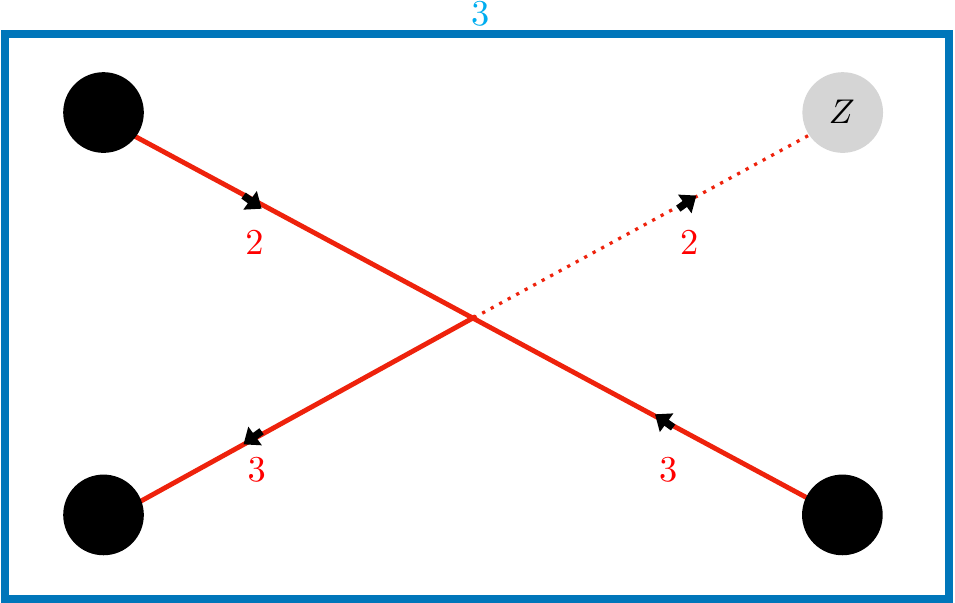}
\caption{\label{fig:addZ} A 3-edge which has been promoted to a 4-edge by the inclusion of an auxiliary boundary vertex $Z$. This allows for flux to be conserved on the hyperedge. Once the HMMF is found, we can ignore all of the flux which connects to $Z$ and then resolve the flux into threads. In this case we have two 3-threads and a single 2-threads. The same process can also be used to describe odd-threads on even hyperedges. }
\end{figure}

To get hypergraphs with non-decomposable odd hyperedges, we will appeal to the hypergraph embedding theorem, which states that a $k$-edge can always be embedded into a $k+1$-edge. For any choice of hypergraph, we construct a new hypergraph which has only even hyperedges. We start by creating a new boundary vertex $Z$ which we add to each odd hyperedge, converting it into an even hyperedge. Since this newly constructed hypergraph is even, it has a HMMF. Now we set each $V_{e_i,Z}=0$. The resulting configuration of fluxes is then our HMMF on the original hypergraph (see figure \ref{fig:addZ}). 

Even though the fluxes on the hyperedges will not be conserved on the pre-embedded odd hyperedges, they are conserved with respect to the higher valence hypergraph into which our hypergraph was embedded into. This complication is only present because of the need to impose orientation to the flow. Once we view the flow as unoriented, it is easy to interpret the flux in terms of individual threads\footnote{This same procedure can also be used to describe odd-threads on even hyperedges.}.

\vspace{\baselineskip}
\noindent These results can be summarized as follows:

\begin{theorem}[Existence of a HMMF]\label{T2}
On every hypergraph there exists a HMMF.
\end{theorem}

\begin{lemma}
On an even hypergraph, there exists a HMMF such that the flux on each hyperedge is conserved. Furthermore, the flux of this HMMF can be decomposed into even-threads.
\end{lemma}
\noindent This only implies the existence of \emph{a} HMMF with these properties. In general one can construct other HMMFs which contain odd-threads.

\subsection{Inequalities}

\paragraph{MMI}
To discuss MMI we work with 4-party pure states and examine hypergraphs with 4 boundary regions $A,B,C,D$. By definition a HMMF on a particular hypergraph is guaranteed to saturate the single party entropies. The subflows of the HMMF can be decomposed into groups of threads connecting a particular combination of boundary regions. The existence of higher party threads affects this decomposition. For example, for $A$ we have
\begin{equation}\label{eq:1partyflow}
    S(A)=V_{A}=N_{AB}+N_{AC} + N_{AD} +N_{ABC}+N_{ABD}+N_{ACD}+N_{ABCD},
\end{equation}
but for $AB$ we have
\begin{equation}\label{eq:2partyflow}
    S(AB)\geq V_{AB}=N_{AC} + N_{AD}+N_{BC} +N_{BD} +N_{ABC}+N_{ABD}+N_{ACD}+N_{BCD}+N_{ABCD}.
\end{equation}
For convenience we define $N_3$ to be the sum of all 3-party threads and $N_4$ all 4-party threads. That is:
\begin{equation}
    N_3=N_{ABC}+N_{ACD}+N_{ABD}+N_{BCD}, \quad N_4=N_{ABCD}.
\end{equation}

We will now show that hypergraphs in general violate MMI. Writing the inequality in terms of entropies we use \eqref{eq:2partyflow} to expand the left hand side in terms of subflows and then use \eqref{eq:1partyflow} to reorganize and write the subflows in terms of the single party entropies
\begin{equation}
    \begin{split}
    S(AB)+S(AC)+S(BC)&\geq V_{AB}+V_{AC}+V_{BC}\\
    &=V_{A}+V_{B}+V_{C}+V_{D}-N_4 \\
    &= S(A)+S(B)+S(C)+S(D)-N_4\\
    I_3&\leq N_4
    \end{split}
\end{equation}
This is exactly MMI except for $N_4$ which is the number of threads in the HMMF which connect all four boundary regions. That is, the presence of four party entanglement causes a hypergraph to violate MMI.

\begin{figure}[H]
\centering
\begin{tabular}{cc}
\includegraphics[width=.5\textwidth,page=13]{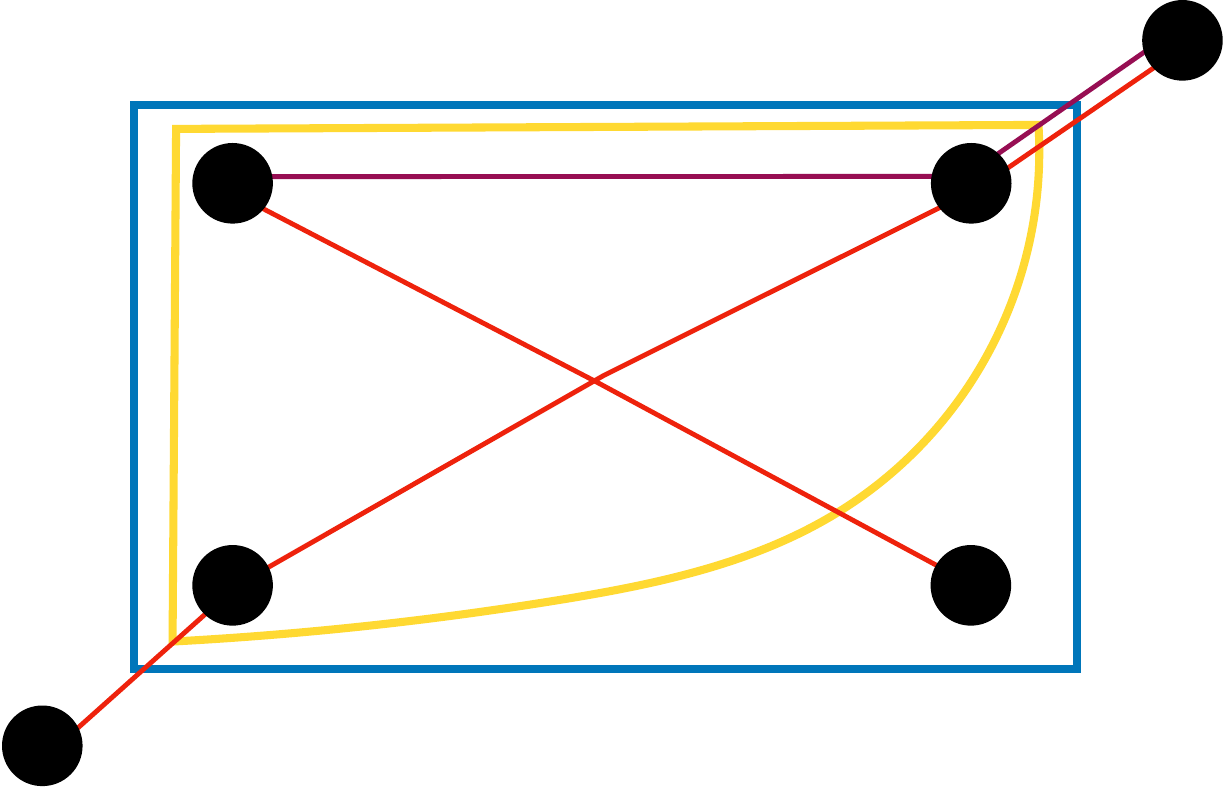} & \includegraphics[width=.5\textwidth,page=14]{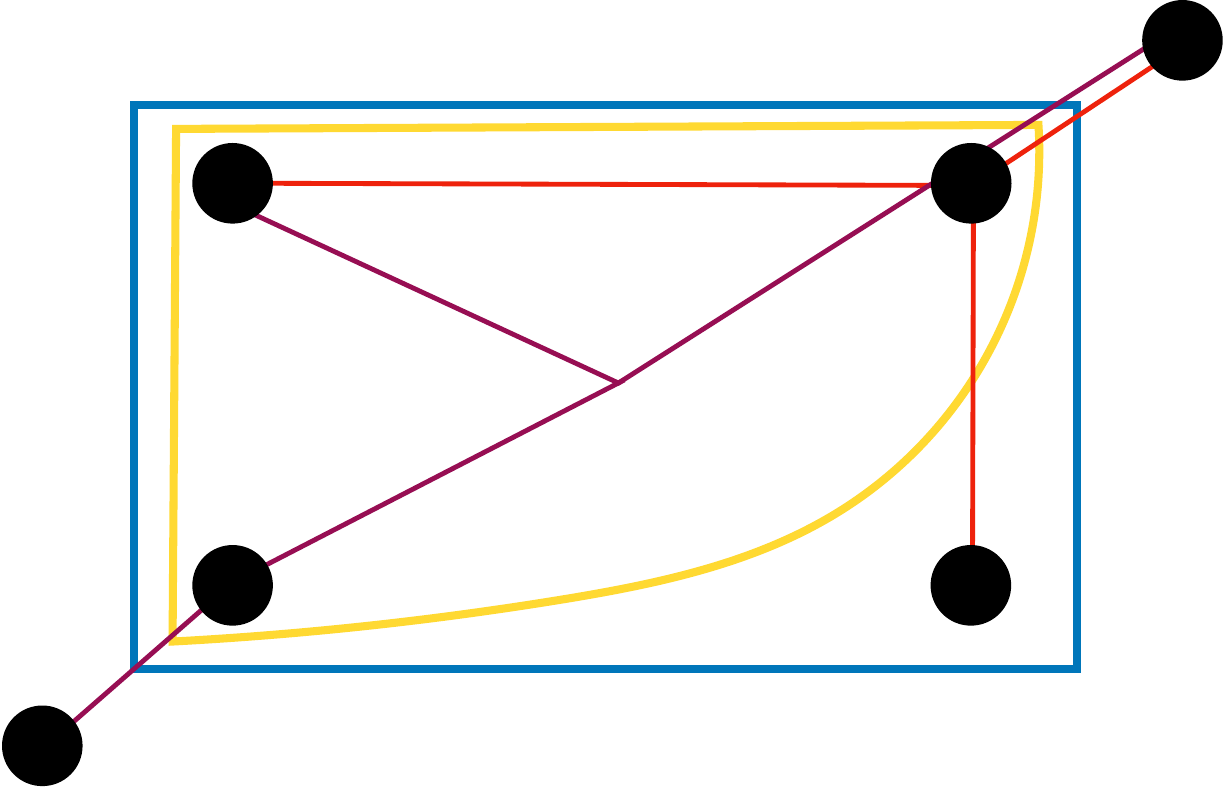}
\end{tabular}
\caption{\label{fig:MMIobstruction} Two HMMF with different thread configurations. Left: $N_{AB}=1$ and $N_{ABCD}=1$ Right: $N_{ABD}=1$ and $N_{ABC}=1$. The existence of this thread configuration without 4-threads means $T_4=0$ and the hypergraph obeys MMI.}
\end{figure}

It is important to note that there will generally exist many HMMF for a hypergraph. Each of these can have a different configuration of subflows. This means the value of $N_4$ depends on the choice of HMMF. For a given hypergraph, we can consider the HMMF where $N_4$ is the smallest, thus producing the tightest bound. This can be thought of as a gauging or an additional constraint, where when searching for a HMMF we choose the one for which $N_4$ is the smallest (see figure \ref{fig:MMIobstruction}). We define schematically
\begin{equation}
    T_4(H) =\min_{\{HMMF\}} N_4
\end{equation}
where $H$ is a hypergraph and the minimization is over the set of HMMF on $H$. As discussed previously, 4-threads act as an avatar of true 4-party entanglement, much like 2-threads are normally regarded as bell pairs. Thus, we can view $T_4$ as an invariant of the hypergraph which measures the amount of 4-party GHZ-like entanglement which can not be transformed into lower party entanglement. With this we can strengthen our bound to
\begin{equation}
   I_{3} \leq T_4.
\end{equation}
Note this also implies that a hypergraph will obey MMI iff $T_4=0$.

\paragraph{Ingleton's inequality}

Because we are now working with a 5-party pure state, we have to adjust our definitions for the subflows. As above we define $N_3,N_4,N_5$ so that we have
\begin{equation}\label{eq:1partyflow5}
    \begin{split}
    S(A)&=V_{A}\\
    &=N_{AB}+N_{AC}+ N_{AD}+N_{AE}\\ &+N_{ABC}+N_{ABD}+N_{ACD}+N_{ABE}+N_{ACE}+N_{ADE}\\
    &+N_4-N_{BCDE}+N_5,
    \end{split}
\end{equation}
but for $AB$ we have
\begin{equation}\label{eq:2partyflow5}
    S(AB)\geq V_{AB}=N_{AC} + N_{AD}+N_{AE}+N_{BC} +N_{BD}+N_{BE}+N_{3}-N_{CDE}+N_4+N_5.
\end{equation}
Note that, using purity of the state, we can relate these to the four and three party entropies respectively.
We can write Ingleton's inequality using entanglement entropies as
\begin{equation}
   S(AB)+ S(AC)+S(AD)+S(BC)+S(BD) \geq S(A)+S(B)+S(CD)+S(ABC)+S(ABD).
\end{equation}
Since the right hand side contains 2-party entropies, the proof method for MMI will not directly follow. We will instead prove a slightly weaker version which assumes that there are no obstructions to the threads computing the 2-party entropies\footnote{It is possible that some further progress could be made by leveraging the bit-thread-based strategy of this proof in the context of the existing full cut-based proof of Ingleton's inequality for hypergraph states; in particular this might lead to a better understanding of the allowed structures of HMMF's on hypergraphs. We will leave this for future work.}. This is more likely for graphs with high valence edges. We proceed by writing
\begin{equation}
\begin{split}
  S(AB)+S(AC)+S(AD)+S(BC)+S(BD)&=V_{AB}+V_{AC}+V_{AD}+V_{BC}+V_{BD}\\
  S(A)+S(B)+S(CD)+S(ABC)+S(ABD)&=V_A+V_B+V_{CD}+V_{DE}+V_{CE}.
\end{split}
\end{equation}
From here we can expand both of these in terms of thread numbers using \eqref{eq:1partyflow5} and \eqref{eq:2partyflow5}. For the left hand side we get
\begin{equation}
    3N_2+5(N_3+N_4+N_5)+N_{AB}+N_{CD}-N_{CE}-N_{DE}-N_{CDE}-N_{BDE}-N_{BCE}-N_{ADE}-N_{ACE}
\end{equation}
while for the right hand side we have
\begin{equation}
    3N_2+4N_3+5(N_4+N_5)-N_{AB}-N_{CD}-N_{CE}-N_{DE}-N_{CDE}-N_{BCDE}-N_{ACDE}.
\end{equation}
We can compare the two sides and simplify to find
\begin{equation}\label{eq:ingv}
    2(N_{AB}+N_{CD})+N_{ABC}+N_{ABD}+N_{ABE}+N_{ACD}+N_{BCD}+N_{CDE}+N_{BCDE}+N_{ACDE}\geq 0
\end{equation}
where the left hand side is explicitly positive. 

It is known that for 2-graphs Ingleton's inequality reduces to MMI and subadditivity. As a non-trivial check we can reproduce this result. To do so, we set all of the thread contributions except the 2-threads to 0. Doing this restricts us to 2-graphs and is implicitly using MMI. \eqref{eq:ingv} becomes
\begin{equation}
    2(N_{AB}+N_{CD})=I(A:B)+I(C:D)\geq 0,
\end{equation}
which is two instances of subadditivity.

It would be interesting to use our framework to prove the full version of Ingleton's inequality, though it is possible that additional machinery may be needed. This is not unlike the difficulties bit threads have in computing higher-party inequalities for 2-graphs and holographic spacetimes \cite{Cui:2018dyq,Headrick:2020gyq}. It is possible that even though bit threads can not always saturate 2-party entropies, particular combinations, those which can be viewed as ``resources", can always be correctly computed. If true this lesson would be a valuable tool for proving these higher party inequalities with bit threads. We state this as the following conjecture:

\begin{figure}[H]
\centering
\includegraphics[width=.5\textwidth,page=15]{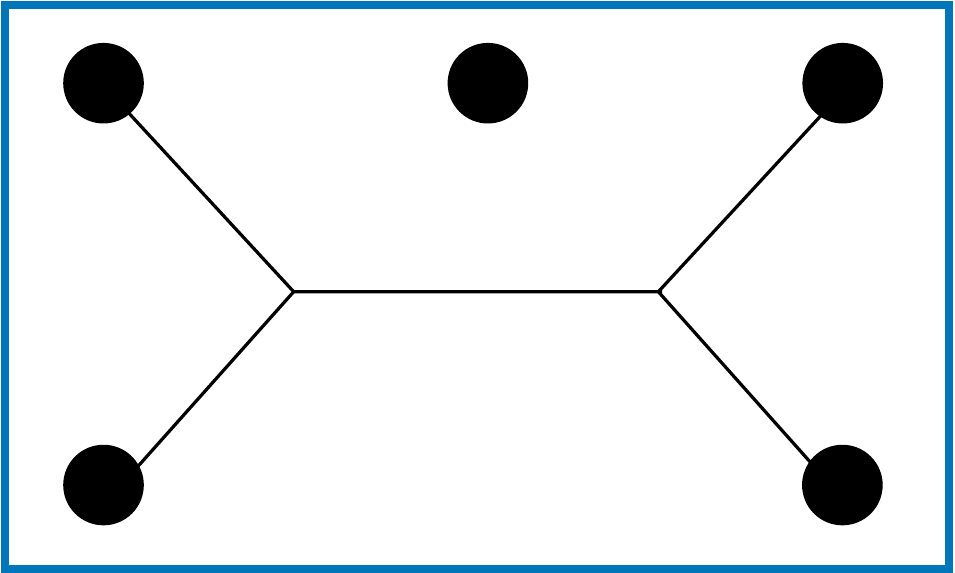}
\caption{\label{fig:5party} A hypergraph with a single 5-edge and several 2-edges. The only HMMF is given by $N_5=N_{AB}=N_{CD}=1$. Note the 2-edges were explicitly chosen such that the HMMF can not simultaneously saturate the 2-party entropies, that is $S(AB)=2\neq V_{AB}=1$. The same is also true for $CD$. Even so we can calculate to find $Ing(S)=Ing(V)=1$.}
\end{figure}

\noindent Let $R(S)$ be defined as
\begin{equation}
    R(S)=S_1-S_2=\sum\alpha_{I}S_{I}-\sum\alpha_{J}S_{J}
\end{equation}
with each $\alpha\geq0$. We define the same linear combination using subflows as
\begin{equation}
    R(V)=V_1-V_2=\sum\alpha_{I}V_{I}-\sum\alpha_{J}V_{J}.
\end{equation}
 This can alternatively be resolved into a linear combination of threads. It is important to note that for a given geometry\footnote{Here we wish to make this more general. We use the term ``geometry" to mean a 2-graphs, hypergraphs, holographic spacetimes or any other system in which one can define entropies from min-cuts and from max multiflows.} while $R(S)$ is constant, the value of $R(V)$ will depend on the choice of multiflow. In this language the saturation of the single party subflows can be written as $S_1\geq V_1$ and $S_2\geq V_2$.

\begin{conjecture}
For a given geometry consider a linear combination of entropies $R(S)$. If $R(V)\geq0$ for all multiflows and there exists a multiflow such that $R(V)\neq0$, then for all max multiflows
\begin{equation}\label{eq:conj}
 R(S)=R(V).
\end{equation}
\end{conjecture}
 
\noindent This is motivated by calculations such as the one in figure \ref{fig:5party}. The conditions are put in place to eliminate inequalities such as MMI for which \eqref{eq:conj} does not hold. If this conjecture is true, it immediately follows that $R(S)\geq 0$. What this would mean is that even though a particular subflow may not match its corresponding entropy, these linear combinations are distinguished in that either the flows or entropies can be used for computation. Note that $Ing(V)\geq0$ and can easily be made non-zero (see \eqref{eq:ingv}) so it meets the conditions for the conjecture.

\paragraph{Cyclic inequalities}
A class of inequalities, known as the cyclic inequalities \cite{Bao_2015}, exists which holographic states satisfy. These are given by
\begin{equation}\label{cyclic}
    \sum^n_{i=1}S(A_i...A_{i+l-1}|A_{i+l}...A_{i+k+l-1})\geq S(A_1...A_n)
\end{equation}
where $S(A|B)=S(AB)-S(B)$ is the conditional entropy. These inequalities are on $n\geq 2k+l$ boundary regions with all indices taken modulo $n$; hypergraphs will generally violate these. To show this we consider a hypergraph with $n+1$ boundary regions and a single $(n+1)$-edge, i.e. the state GHZ$_{n+1}$. For this hypergraph the HMMF is a single $(n+1)$-thread so that $N_{n+1}=1$ and all others are zero. Expanding \eqref{cyclic} we have $n$ entropies on the left, but $n+1$ on the right which gives
\begin{equation}
   n N_{n+1}\ngeq  (n+1)N_{n+1}.
\end{equation}
In the language above, all of these inequalities have hypergraphs such that $R(V)\leq0$.

\subsection{Connections with holography}
The question naturally arises if hypergraphs and hypercuts can be interpreted in holography in a similar manner as 2-graphs and cuts. As stated previously hypergraphs include states which do not satisfy MMI, meaning they can not be represented by a state on the boundary CFT which is dual to a holographic spacetime. In other words such states are ``non-geometric".

\begin{figure}[H]
\centering
\begin{tabular}{cc}
\includegraphics[width=.4\textwidth,page=16]{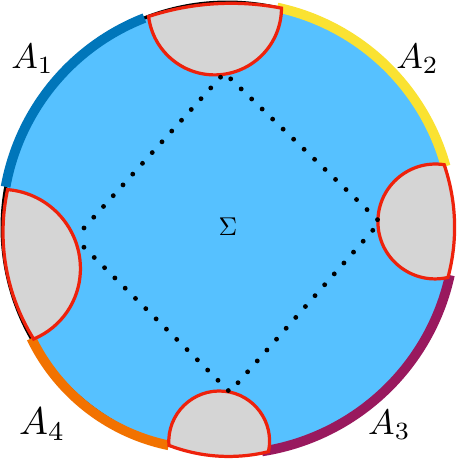} & \includegraphics[width=.4\textwidth,page=17]{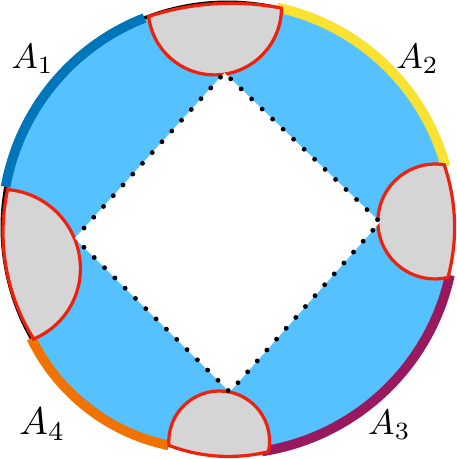}
\end{tabular}
\caption{\label{fig:EWCS}Left: Four regions, their homology region, and multipartite entanglement wedge cross section, $\Sigma$. Right: After performing the ``hypercut" the four regions are now disconnected.}
\end{figure}

Regardless of this, there is a surface which can be constructed from AdS which seems to mimic the behavior of a hypercut in that if we perform the cut, it separates all of the boundary regions from one another simultaneously (see figure \ref{fig:EWCS}). This surface is the multipartite entanglement wedge cross section \cite{Umemoto_2018,Umemoto2_2018,Nguyen_2018,Bao_2018,Bao2_2019} which here we call $\Sigma$. For this argument we focus on the case of 4 non-overlapping boundary regions of AdS. To construct $\Sigma$, we start by calculating the surface $m(ABCD)$, which is the minimal surface homologous to $ABCD$ and whose area calculates $S(ABCD)$. This surface, along with the boundary regions, defines for us the homology region of $ABCD$.\footnote{In many papers the term entanglement wedge is used to refer to the homology region.} Now we calculate the minimal surface homologous to $ABCD$ relative to $m(ABCD)$ with the added constraint that it must form a single connected surface. The result of this optimization is $\Sigma$.

We now notice, that were we to cut $\Sigma$, the homology region would split into 5 separate pieces: one for each boundary and one internal region. In this sense, much like our GHZ state, this single cut has eliminated the shared entanglement.

It is possible to construct purifying manifolds out of several copies of the original homology region. On these manifolds $\Sigma$ can be viewed as a minimal surface in some homology class whose area is the multipartite reflected entropy \cite{Bao3_2019, harper2020multipartite,Chu_2020}. Bit thread constructions exist for both $\Sigma$ and the corresponding surface on the purification \cite{Harper_2019,harper2020multipartite}. Using these tools it is possible that entropy inequalities may be derived on the purification so that if the surfaces and flows are projected down to the homology region they would show $\Sigma$ violates an MMI-like inequality similar to that of GHZ-like entanglement. We leave this and other connections to holography to future work.

\acknowledgments
The work of J.H. is supported by the Simons Foundation through \emph{It from Qubit: Simons Collaboration on Quantum Fields, Gravity, and Information}. N.B. is supported by the National Science Foundation under grant number 82248-13067-44-PHPXH, by the Department of Energy under grant number DE-SC0019380, and by the Computational Science Initiative at Brookhaven National Laboratory. J.H would like to thank Matt Headrick for discussions and comments. J.H would also like to thank Catherine Tate and Josh Levin for reading early versions of this paper.

\bibliographystyle{JHEP}
\bibliography{BT_Hyper}
\end{document}